\documentclass[aps,floats,twocolumn,epsf,prb,showpacs]{revtex4}
\usepackage{graphicx}

\begin{document}

\title{Comparing pertinent effects of antiferromagnetic fluctuations in the
two and three dimensional Hubbard model}
\author{A. A. Katanin$^{a,b},$ A. Toschi$^{a,c}$, and K. Held$^c$}
\affiliation{$^a$Max-Planck-Institut f\"ur Festk\"orperforschung, 70569 Stuttgart, Germany%
\\
$^b$Institute of Metal Physics, 620044 Ekaterinburg, Russia \\
$^c$Institute of Solid State Physics, Vienna University of Technology, 1040
Vienna, Austria}
\date{Version 8, \today }

\begin{abstract}
We use the dynamical vertex approximation (D$\Gamma$A) with a Moriyaesque $%
\lambda$ correction for studying the impact of antiferromagnetic
fluctuations on the spectral function of the Hubbard model in two and three
dimensions. Our results show the suppression of the quasiparticle weight in
three dimensions and dramatically stronger impact of spin fluctuations in
two dimensions where the pseudogap is formed at low enough temperatures.
Even in the presence of the Hubbard subbands, the origin of the pseudogap at
weak-to-intermediate coupling is in the splitting of the quasiparticle peak.
At stronger coupling (closer to the insulating phase) the splitting of
Hubbard subbands is expected instead. The $\mathbf{k}$-dependence of the
self energy appears to be also much more pronounced in two dimensions as can
be observed in the $\mathbf{k}$-resolved D$\Gamma$A spectra, experimentally
accessible by angular resolved photoemission spectroscopy in layered
correlated systems.
\end{abstract}

\maketitle

\let\n=\nu \let\o =\omega \let\s=\sigma


\section{Introduction}

Since its formulation, \cite{Hubbard} the Hubbard model served as a minimal
model for electronic correlations. Due to the complexity of electronic
correlations, solving this model is however only possible in dimension $d=1$
(exactly via the Bethe Ansatz\cite{Bethe}) and in the limit $d=\infty $\cite%
{DMFT,DMFT2,DMFTREV} (where the mapping\cite{DMFT2} onto an Anderson
impurity model allows for an accurate numerical solution \cite{DMFTREV,Bulla}%
). Of physical interest are however strongly correlated systems in $%
d=3$, for modeling the Mott-Hubbard transition \cite{MH} and
(anti-)ferromagnetism \cite{AFM,Hubbard,FM}, and in $d=2$ for describing the
cuprates \cite{Dagotto}, where the role of the antiferromagnetic
fluctuations in developing pseudogap structures and superconductivity are at
the center of attention.

The aim of this paper is to study the difference between the effect of
antiferromagnetic fluctuations on the electronic properties in $d=2$ and $%
d=3 $. For weak coupling (small Coulomb interaction $U$), the perturbation
theory, and its extensions, e.g. the fluctuation-exchange approximation
(FLEX)\cite{FLEX}, the two-particle self-consistent approximation (TPSC)\cite%
{TPSC}, and the functional renormalization group \cite{fRG} are suitable
methods for this purpose. In $d=3$ antiferromagnetic fluctuations produce
only quantitative changes of electronic spectrum, although the particle-hole
excitations enhance the quasiparticle scattering rate when the temperature $T
$ is approaching the N\'{e}el temperature. In $d=2$ there are divergences in
the self-energy diagrams and the abovementioned approximations predict
pseudogap structures in the self-energy in the weak-coupling regime \cite%
{FlexSE,TPSCSE,fRGSE}. 
These techniques are however not applicable at stronger coupling, since they
do not describe strong quasiparticle renormalization due to the Mott physics.

Since we are interested in intermediate-to-strong electronic correlations,
we need to take a different approach. Starting point is the by-now widely
employed dynamical mean-field theory (DMFT).\cite{DMFT,DMFT2,DMFTREV} This
method becomes exact \cite{DMFT} for $d\rightarrow \infty $, and yields a
major part of the electronic correlations, i.e., the local correlations.
However, any non-local correlations are neglected and hence DMFT does not
differentiate between the Hubbard model in two- and three dimensions. More
precisely, only differences stemming from different shapes of the density of
states (DOS) are taken into account, not those resulting, e.g., from
antiferromagnetic correlations since these correlations are by nature
non-local.

Hitherto, the focus of DMFT extensions has been on \emph{short-range}
correlations within a (finite) cluster instead of the single DMFT impurity
site. These cluster extensions of DMFT \cite{clusterDMFT} have been used for
describing pseudogaps and superconductivity in the two-dimensional Hubbard
model. Due to numerical limitations, the inclusion of important \emph{%
long-range} correlations and the application of this method in three
dimensions or realistic multi-orbital calculations is however not possible,
except for very small clusters with $\mathcal{O}(2 \div 4)$ sites. Also the $%
1/d$ expansion of DMFT \cite{Schiller} is restricted to \emph{short-range}
correlations, as is a recent perturbative extension. \cite{Tokar07a}

Hence, for including \emph{long-range} correlations, the focus of the
methodological development has shifted recently to diagrammatic extensions
of DMFT such as the dynamical vertex approximation (D$\Gamma $A) \cite%
{DGA1,DGA2,Kusunose,Monien} and the dual fermion approach by Rubtsov \emph{%
et al.} \cite{DualFermion} Even before, Kuchinskii \textsl{et al.} \cite%
{Sadovskii05} combined the local DMFT self energy with the non-local
contributions to self energy of the spin-fermion model, and included
long-range correlations this way. Their procedure, however, does not rely on
a rigorous diagrammatic derivation.

To include long-range fluctuations in a diagrammatic way D$\Gamma $A
considers the local vertex instead of the bare interaction. It includes DMFT
but also long-range correlations beyond. Our understanding of the physics
associated with such long-range correlation is typically based on ladder
diagrams, which are considered, e.g. by the abovementioned TPSC and FLEX
approximations. For example, the ladder diagrams in the particle-hole
channel yield antiferromagnetic fluctuations in the paramagnetic phase
(paramagnons) and (anti-)ferromagnons in the ordered state. It is natural to
suppose that the contribution of the corresponding fluctuations in the
intermediate coupling regime can be described by the same kind of diagrams
albeit with the \textit{renormalized} vertices. In D$\Gamma$A the local
(frequency dependent) vertex is considered instead of the bare interaction.
Therefore, this method reproduces the results of the weak-coupling
approaches at small $U$ but can treat spatial correlations also at
intermediate coupling. Hence, D$\Gamma $A is well suited for studying
antiferromagnetic fluctuations in strongly correlated systems both for $d=2$
and $d=3$.

The paper is organized as follows: In Section \ref{Sec:DGA} we reiterate the
D$\Gamma $A approach in a formulation with the three-point (instead of the
four-point) vertex functions which allows for a connection to the spin
fermion model in Section \ref{Sec:SF} and for the analytical considerations
on the D$\Gamma $A self energy in Section \ref{Sec:approx}. In Section \ref%
{Sec:lambda}, we introduce a Moriyaesque $\lambda $ correction to the
susceptibility to describe correctly the two-dimensional case. Results for
three dimensions are presented in Section \ref{Sec:3D} and compared to
those in two dimensions in Section \ref{Sec:2D}. Special emphasis to angular
resolved spectra is given in Section \ref{Sec:ARPES} before we give a brief
summary in Section \ref{Sec:Conclusion}.

\section{Dynamical Vertex Approximation}

\label{Sec:DGA}

Starting point of our considerations is the Hubbard model on a square or
cubic lattice 
\begin{equation}
H=-t\sum_{\langle ij\rangle \sigma }c_{i\sigma }^{\dagger }c_{j\sigma
}+U\sum_{i}n_{i\uparrow }n_{i\downarrow }  \label{H}
\end{equation}%
where $t$ denotes the hopping amplitude between nearest-neighbors, $U$ the
Coulomb interaction, $c_{i\sigma }^{\dagger }$($c_{i\sigma }$) creates
(annihilates) an electron with spin $\sigma $ on site $i$; $n_{i\sigma
}\!=\!c_{i\sigma }^{\dagger }c_{i\sigma }$. In the following, we restrict
ourselves to the paramagnetic phase with $n=1$ electrons/site at a finite
temperature $T$. 

The D$\Gamma $A result for the self-energy of the model (\ref{H}) was
derived in Ref. \onlinecite{DGA1}, see Eq. (16). For the purpose of the
present paper this result for the self-energy can be written in the form 

\begin{eqnarray}
\Sigma _{\mathbf{k},\nu } &=&\frac{1}{2}{Un}+\frac{1}{2}TU\sum\limits_{\nu
^{\prime }\nu ^{\prime \prime }\omega ,\mathbf{q}}\left[ 3\chi _{s,\mathbf{q}%
}^{\nu ^{\prime }\nu ^{\prime \prime }\omega }\Gamma _{s,\text{ir}}^{\nu
^{\prime \prime }\nu \omega }-\chi _{c,\mathbf{q}}^{\nu ^{\prime }\nu
^{\prime \prime }\omega }\Gamma _{c,\text{ir}}^{\nu ^{\prime \prime }\nu
\omega }\right.   \nonumber \\
&&\left. +\chi _{0\mathbf{q}\omega }^{\nu ^{\prime }}(\Gamma _{c,\text{loc}%
}^{\nu \nu ^{\prime }\omega }-\Gamma _{s,\text{loc}}^{\nu \nu ^{\prime
}\omega })\right] G_{\mathbf{k+q},\nu +\omega },  \label{SE}
\end{eqnarray}%

where the non-local spin (s) and charge (c) susceptibilities%
\begin{equation}
\chi _{s(c),\mathbf{q}}^{\nu \nu ^{\prime }\omega }=[(\chi _{0\mathbf{q}%
\omega }^{\nu ^{\prime }})^{-1}\delta _{\nu \nu ^{^{\prime }}}-\Gamma _{s(c),%
\text{ir}}^{\nu \nu ^{\prime }\omega }]^{-1}
\end{equation}%
can be expressed in terms of the particle-hole bubble $\chi _{0\mathbf{q}%
\omega }^{\nu ^{\prime }}=-T\sum_{\mathbf{k}}G_{\mathbf{k},\nu ^{\prime }}G_{%
\mathbf{k}+\mathbf{q},\nu ^{\prime }+\omega }$, $G_{\mathbf{k},\nu }=[i\nu
-\epsilon _{\mathbf{k}}+\mu -\Sigma _{\text{loc}}(\nu )]^{-1}$ is the Green
function, and $\Sigma _{\text{loc}}(\nu )$ the local self-energy. The spin
(charge) irreducible local vertices $\Gamma _{s(c),\text{ir}}^{\nu \nu
^{\prime }\omega }$ are determined from the corresponding local problem\cite%
{DGA1}.

\begin{figure}[tb]
\includegraphics[width=8cm]{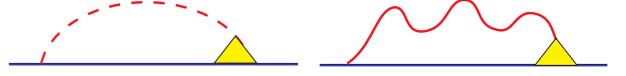}
\caption{(Color online) Graphical representation of the contribution of bare
Coulomb interaction (a) and spin (charge) fluctuations (b) to the
self-energy in the D$\Gamma$A approach, Eq. (\protect\ref{SE0}). Solid lines
correspond to the electronic Green function $G_{\mathbf{k},\protect\nu} $,
dashed line to the bare Hubbard interaction $U$, wiggly lines - to the spin
(charge) susceptibility $\protect\chi^{\mathrm{s(c)}}_{\mathbf{q},\protect%
\omega}$; the triangle corresponds to the interaction vertex $\protect\gamma%
^{\protect\nu,\protect\omega}_{\mathrm{s(c)},\mathbf{q}}$. }
\label{Fig:1}
\end{figure}

The result (\ref{SE}) accounts for the contribution of ladder diagrams to
the self-energy in the two particle-hole channels. Following Edwards and
Hertz \cite{HertzEdw} it is convenient to pick out parts of these ladders,
which are separated by the bare on-site Coulomb interaction $U.$ This is
achieved by considering the quantities 
\begin{eqnarray}
\Phi _{s(c),q}^{\nu \nu ^{\prime }\omega } &=&[(\chi _{0\mathbf{q}\omega
}^{\nu ^{\prime }})^{-1}\delta _{\nu \nu ^{^{\prime }}}-\Gamma _{s(c),\text{%
ir}}^{\nu \nu ^{\prime }\omega }\pm U]^{-1},  \label{Fi} \\
\phi _{\mathbf{q},\omega }^{s(c)} &=&\sum\limits_{\nu \nu ^{\prime }}\Phi
_{s(c),\mathbf{q}}^{\nu \nu ^{\prime }\omega }  \nonumber
\end{eqnarray}%
such that $\chi _{s(c),\mathbf{q}}^{\nu \nu ^{\prime }\omega }=\{[\Phi
_{s(c),q}^{\nu \nu ^{\prime }\omega }]^{-1}\mp U\}^{-1}$ with the upper
(lower) sign for the spin (charge) susceptibility. The nonlocal spin
(charge) susceptibility is then given by%
\begin{equation}
\chi _{\mathbf{q}\omega }^{s(c)}=\sum\limits_{\nu \nu ^{\prime }}\chi _{s(c),%
\mathbf{q}}^{\nu \nu ^{\prime }\omega }=[(\phi _{\mathbf{q,}\omega
}^{s(c)})^{-1}\mp U]^{-1}.  \label{hi}
\end{equation}%
and therefore $\phi _{\mathbf{q,}\omega }^{s(c)}$ provided to be a
particle-hole irreducible susceptibility in the spin (charge) channel.
Introducing, similar to Ref.\ \onlinecite{HertzEdw}, the corresponding
three-point vertex $\gamma _{s(c),\mathbf{q}}^{\nu \omega }$ of electron
interaction with charge (spin) fluctuations,%
\begin{equation}
\gamma _{s(c),\mathbf{q}}^{\nu \omega }=(\chi _{0\mathbf{q}\omega }^{\nu
})^{-1}\sum\limits_{\nu ^{\prime }}\Phi _{s(c),\mathbf{q}}^{\nu \nu ^{\prime
}\omega },  \label{gamma}
\end{equation}%
the irreducible susceptibility $\phi _{\mathbf{q,}\omega }^{s(c)}$ can be
represented as 
\begin{equation}
\phi _{\mathbf{q,}\omega }^{s(c)}=\sum\limits_{\nu }\gamma _{s(c),\mathbf{q}%
}^{\nu \omega }\chi _{0\mathbf{q}\omega }^{\nu }  \label{fi}
\end{equation}%
In these notations, the result (\ref{SE}) can then be rewritten identically
as%
\begin{eqnarray}
\Sigma _{\mathbf{k},\nu } &=&\frac{1}{2}{Un}+\frac{1}{2}TU\sum\limits_{%
\omega ,\mathbf{q}}\left[ 3\gamma _{s,\mathbf{q}}^{\nu \omega }-\gamma _{c,%
\mathbf{q}}^{\nu \omega }-2\right.  \nonumber \\
&&+3U\gamma _{s,\mathbf{q}}^{\nu \omega }\chi _{\mathbf{q}\omega
}^{s}+U\gamma _{c,\mathbf{q}}^{\nu \omega }\chi _{\mathbf{q}\omega }^{c} 
\nonumber \\
&& +\sum\limits_{\nu ^{\prime }} \left. \chi _{0\mathbf{q}\omega }^{\nu
^{\prime }}(\Gamma _{c,\text{loc}}^{\nu \nu ^{\prime }\omega }-\Gamma _{s,%
\text{loc}}^{\nu \nu ^{\prime }\omega })\right] G_{\mathbf{k+q},\nu +\omega }
\label{SE0}
\end{eqnarray}%
The first three terms in the square brackets correspond to the interaction
of electrons via Hubbard on-site Coulomb interaction (without forming
ph-bubbles, Fig. \ref{Fig:1}a), the next two terms correspond to electron
interactions via charge- and spin-fluctuations (Fig. \ref{Fig:1}b), the last
term subtracts double counted local contribution.

\section{Relation to spin-fermion models}

\label{Sec:SF} The contributions of bare Coulomb interaction and charge
(spin) fluctuations to the self-energy (\ref{SE0}) can be also obtained from
the fermion-boson model with generating functional%
\begin{eqnarray}
&&Z%
\begin{array}{c}
=%
\end{array}%
\int D[c_{k\sigma }^{\dagger },c_{k\sigma }]D\mathbf{S}_{\mathbf{q},\omega
}D\rho _{\mathbf{q},\omega }\exp \{-\mathcal{L}[\mathbf{S},\rho ,c]\} 
\nonumber \\
&&\ \mathcal{L}[\mathbf{S},\rho ,c]%
\begin{array}{c}
=%
\end{array}%
\sum\limits_{\mathbf{k},\nu, \sigma }(i\nu _{n}-\varepsilon _{\mathbf{k}%
})c_{k\sigma }^{\dagger }c_{k\sigma }  \label{sf} \\
&&\ \ \ +U\sum\limits_{\mathbf{q},\omega }(\rho _{q\omega }\rho _{-q,-\omega
}+\mathbf{S}_{q\omega }\mathbf{S}_{-q,-\omega })  \nonumber \\
&&\ \ \ +U\sum\limits_{\mathbf{k,q,}\nu ,\omega, \sigma,\sigma^{\prime}
}(\gamma _{s,\mathbf{q}}^{\nu \omega })^{1/2}c_{\mathbf{k},\nu ,\sigma
}^{\dagger }\mbox {\boldmath $\sigma
$}_{\sigma \sigma ^{\prime }}c_{\mathbf{k+q},\nu +\omega ,\sigma ^{\prime }}%
\mathbf{S}_{\mathbf{q},\omega }  \nonumber \\
&&\ \ \ +iU\sum\limits_{\mathbf{k,q,}\nu ,\omega }(\gamma _{c,\mathbf{q}%
}^{\nu \omega })^{1/2}c_{\mathbf{k},\nu ,\sigma }^{\dagger }c_{\mathbf{k+q}%
,\nu +\omega ,\sigma }\rho _{\mathbf{q},\omega }  \nonumber
\end{eqnarray}%
where $\gamma _{c(s),\mathbf{q}}^{\nu \omega }$ is determined in the present
approach according to the Eq. (\ref{gamma}) and $\mbox {\boldmath $\sigma
$}_{\sigma \sigma ^{\prime }}$ are the Pauli matrices. The model (\ref{sf})
is similar to that derived from the Hubbard model via Hubbard-Stratonovich
transformation\cite{Hertz}, but it is explicitly spin symmetric and contains
the non-local frequency dependent vertices $\gamma _{c(s),\mathbf{q}}^{\nu
\omega },$ which account for the local- and short range-nonlocal
fluctuations.

\begin{figure}[tb]
\includegraphics[width=8cm]{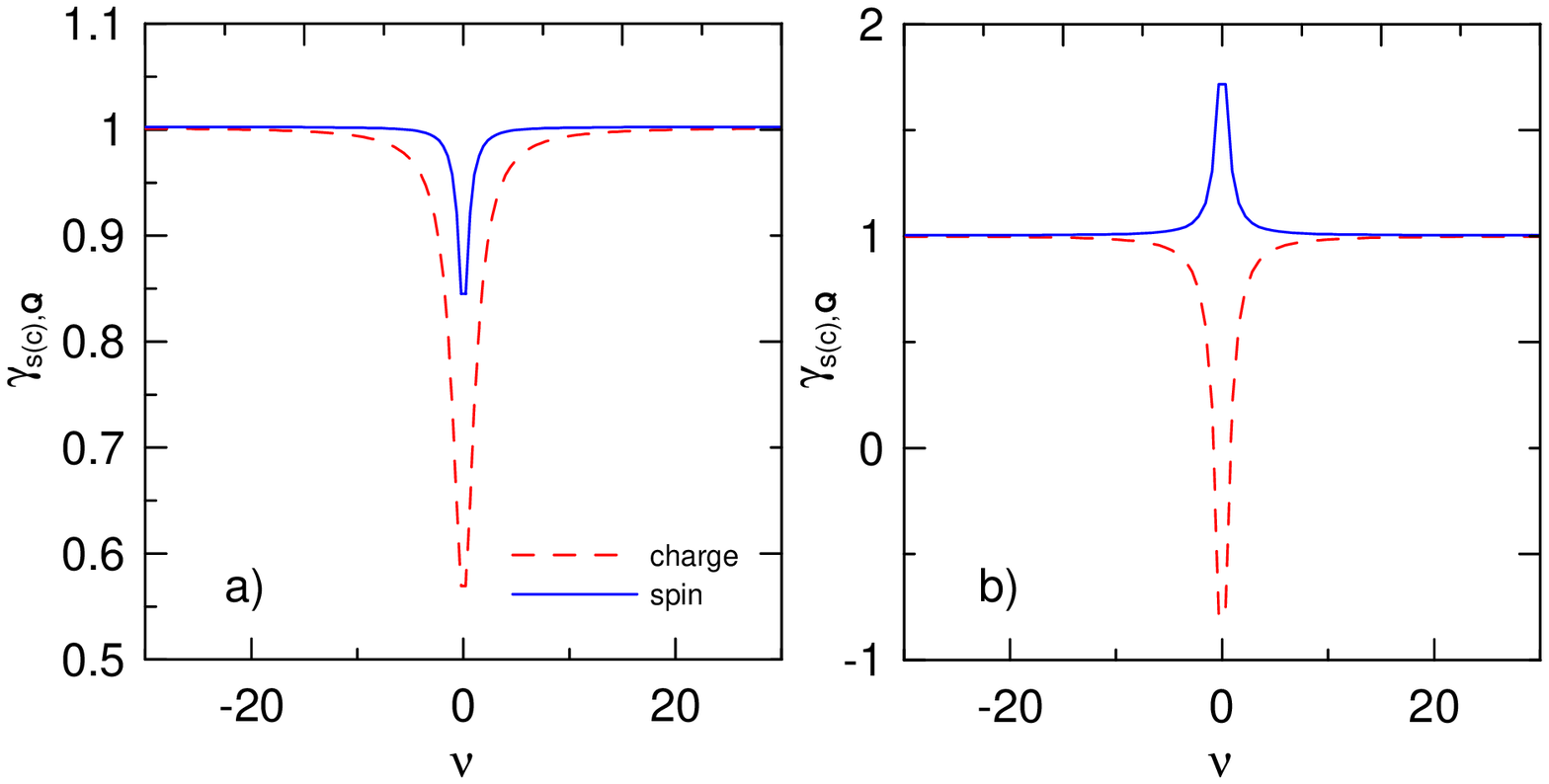}
\caption{(Color online) Frequency dependence of the spin and charge
three-point vertex Eq.\ (\protect\ref{gamma}) at $U=1$, $\protect\beta=1/T=15
$ (left) and $U=2$, $\protect\beta =10$ (right), $\protect\omega =0$, at
the antiferromagnetic wave vector $\mathbf{Q}=(\protect\pi ,\protect\pi )$. All energies are in units of half the effective bandwidth $%
D\equiv 4t$.}
\label{Fig:2}
\end{figure}

Contrary to the earlier paramagnon theories\cite{Moriya} and the
spin-fermion model\cite{SpFerm,SpFerm1}, where $\gamma _{s,\mathbf{q}}^{\nu
\omega }=1$ and charge fluctuations are omitted ($\gamma _{c,\mathbf{q}%
}^{\nu \omega }=0$), we have $\gamma _{s(c),\mathbf{q}}^{\nu \omega }\neq 0$
and $\neq 1$. The frequency dependence of the vertices $\gamma _{s(c),%
\mathbf{Q}}^{\nu 0} $ calculated in the present approach for two dimensions
with $\mathbf{Q}=(\pi,\pi)$ is shown in Fig. \ref{Fig:2} (in the three
dimensional case we observe qualitatively similar behavior). One can see,
that both charge- and spin vertices have a strong frequency dependence and
approach unity only in the high-frequency limit. While in the weak-coupling
regime $U=D\equiv 4t$ both vertices are suppressed at small frequencies
[which is the consequence of the particle-particle (Kanamori) screening],
closer to the DMFT Mott transition (at $U=2D\equiv 8t$) the spin vertex at
small frequencies is \textit{enhanced}. This behavior is similar to that
observed in Ref.\ \onlinecite{DGA1} for the three-frequency (four-point)
vertex in the three dimensional case.

Hence, the spin-fermion theory, which was heuristically added to the DMFT
self-energy before, is included in a more systematic and consistent way in D$%
\Gamma $A, which also accounts for the corrections to the
electron-paramagnon vertex. The susceptibility $\chi _{q,\omega }^{s}$ which
is determined phenomenologically in the spin-fermion model is obtained in
our approach by dressing the bare propagator $1/U$ of charge- and spin
fields by particle-hole bubbles, which reproduces the results (\ref{hi}) and
(\ref{fi}) of the previous Section.

Using the model (\ref{sf}) one can also calculate the leading order
non-local correction to the three-point vertices due to fermion-boson
interaction, 
\begin{eqnarray}
\widetilde{\gamma }_{s,\mathbf{k},\mathbf{q}}^{\nu \omega } &=&\gamma _{s,%
\mathbf{q}}^{\nu \omega }+\frac{1}{2}TU\sum\limits_{\omega _{1},\mathbf{q}%
_{1}}\gamma _{s,\mathbf{q}}^{\nu +\omega _{1},\omega }\left[ 2-\gamma _{s,%
\mathbf{q}_{1}}^{\nu \omega _{1}}-\gamma _{c,\mathbf{q}_{1}}^{\nu \omega
_{1}}\right.  \nonumber \\
&&\left. -U\gamma _{s,\mathbf{q}_{1}}^{\nu \omega _{1}}\chi _{\mathbf{q}%
_{1},\omega _{1}}^{s}+U\gamma _{c,\mathbf{q}_{1}}^{\nu \omega _{1}}\chi _{%
\mathbf{q}_{1}\omega _{1}}^{c}\right] G_{\mathbf{k+q}_{1},\nu +\omega _{1}} 
\nonumber \\
&&\times G_{\mathbf{k+q}_{1}+\mathbf{q},\nu +\omega _{1}+\omega }^{{}}-\text{%
loc,} \\
\widetilde{\gamma }_{c,\mathbf{k},\mathbf{q}}^{\nu \omega } &=&\gamma _{s,%
\mathbf{q}}^{\nu \omega }+\frac{1}{2}TU\sum\limits_{\omega _{1},\mathbf{q}%
_{1}}\gamma _{s,\mathbf{q}}^{\nu +\omega _{1},\omega }\left[ 3\gamma _{s,%
\mathbf{q}_{1}}^{\nu \omega _{1}}-\gamma _{c,\mathbf{q}_{1}}^{\nu \omega
_{1}}-2\right.  \nonumber \\
&&\left. +3U\gamma _{s,\mathbf{q}_{1}}^{\nu \omega _{1}}\chi _{\mathbf{q}%
_{1},\omega _{1}}^{s}+U\gamma _{c,\mathbf{q}_{1}}^{\nu \omega _{1}}\chi _{%
\mathbf{q}_{1}\omega _{1}}^{c}\right] G_{\mathbf{k+q}_{1},\nu +\omega _{1}} 
\nonumber \\
&&\times G_{\mathbf{k+q}_{1}+\mathbf{q},\nu +\omega _{1}+\omega }^{{}}-\text{%
loc,}
\end{eqnarray}%
where loc stands for the subtraction of the local terms already included in $%
\gamma _{s,\mathbf{q}}^{\nu \omega }$. The non-local corrections to the
self-energy and vertex can be then treated self-consistently by substituting
them into Eq. (\ref{fi}). This provides an alternative simpler way of 
self-consistent treatment instead of the more complicated parquet approach
discussed in Ref. \onlinecite{DGA1}. 
 An even simpler way to go beyond a non-self consistent treatment of the D%
$\Gamma$A equations is considered in Sect. V.

\section{Analytic approximation for the D$\Gamma $A self energy}

\label{Sec:approx}

Similarly to the weak-coupling approach \cite{TPSC}, in the two dimensional
case the self-energy can be obtained approximately analytically. In this
case the susceptibility $\chi _{\mathbf{q}\omega }^{s}$ is strongly enhanced
at $\omega _{n}=0$ and $\mathbf{q}\approx \mathbf{Q}=(\pi ,\pi )$, and can
be represented in the form%
\begin{equation}
\chi _{\mathbf{q}0}^{s}=\frac{A}{(\mathbf{q}-\mathbf{Q})^{2}+\xi ^{-2}}
\label{hiq}
\end{equation}%
where $\xi ^{-2}=A/(1-U\phi _{\mathbf{Q}0}^{s})$ with $A=(\nabla ^{2}\phi _{%
\mathbf{q}0}^{s})_{\mathbf{q}=\mathbf{Q}}$ being the (squared) inverse spin
fluctuation correlation length. Since the corresponding momentum sum in the
Eq. (\ref{SE0}) over $\mathbf{q}$ is logarithmically divergent at $\xi
\rightarrow \infty $, we can approximately retain ourselves to only the 
zero bosonic Matsubara frequency term in the spin-fluctuation contribution
and put $\mathbf{q}\approx \mathbf{Q}$ in all the factors except $\chi _{%
\mathbf{q}0}^{s}$ to obtain%
\begin{equation}
\Sigma _{\mathbf{k},\nu }\simeq \Sigma _{\text{loc}}(\nu )+\Delta ^{2}\gamma
_{s,\mathbf{Q}}^{\nu ,0}G_{\mathbf{k}+\mathbf{Q},\nu }  \label{SEAprox}
\end{equation}%
where $\Delta ^{2}=(3TU^{2}/2)\sum\limits_{\mathbf{q}}\chi _{\mathbf{q}%
,0}^{s}.$

To study the frequency dependence of the self-energy (\ref{SEAprox})
qualitatively, we first consider $\gamma _{s,\mathbf{Q}}^{\nu ,0}=1$ and
choose the local self-energy in the form (see, e.g. Ref.\ \onlinecite{Biczuk}%
) 
\begin{equation}
\Sigma _{\text{loc}}(\nu )=(1-\kappa )(\Delta _{\text{loc}}^{2}/4)/(\nu
-\Delta _{\text{loc}}^{2}\kappa /(4\nu ))  \label{Eq:SigmaKappa}
\end{equation}%
where $\Delta _{\text{loc}}\simeq U$ is the size of the Hubbard gap and $%
\kappa $ measures the relative weight of the quasiparticle peak (QP) with
respect to the Hubbard subbands ($\kappa =0$ at the Mott transition and $%
\kappa =1$ for $U\rightarrow 0$). The Eq. (\ref{Eq:SigmaKappa}) allows to
reproduce the three-peak structure of the self-energy, observed in the
numerical solution of the single-impurity Anderson model, supplemented by
the DMFT self-consistent condition.

\begin{figure}[tb]
\centerline{
\hspace{.2cm}\includegraphics[width=10.2cm]{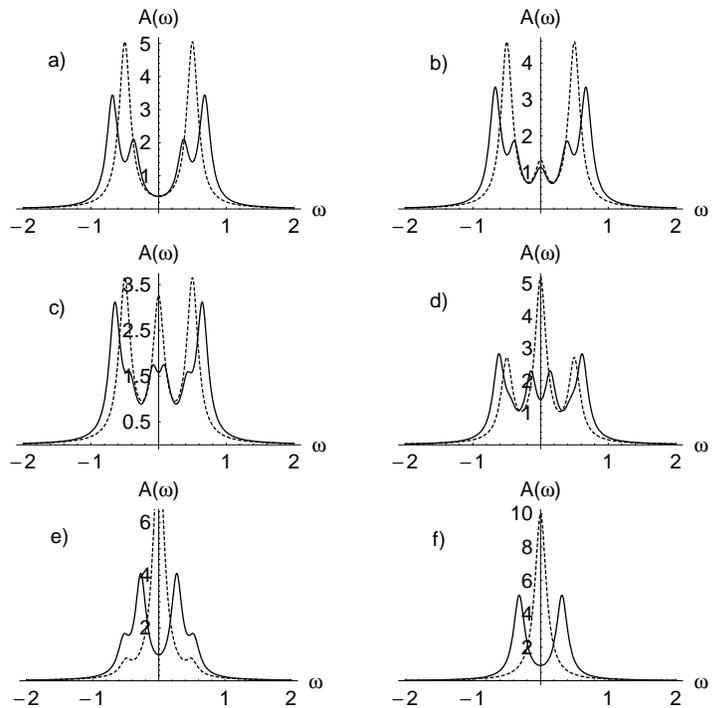}}
\caption{(Color online) The spectral functions in $d=2$ as obtained from the
approximate self-energies including local (dashed lines, Eq. (\protect\ref%
{Eq:SigmaKappa})) and non-local (solid lines, Eq. (\protect\ref{SEAprox}))
fluctuations for $\protect\kappa =0$ (a), 0.1 (b), 0.3 (c), 0.5 (d), 0.9
(e), and 1.0 (f). }
\label{Fig:3}
\end{figure}

The evolution of the spectral properties calculated with the self-energies (%
\ref{SEAprox}) and (\ref{Eq:SigmaKappa}) with changing $\kappa $ for $\Delta
_{\text{loc}}=1$ and $\Delta =0.1$ is shown in Fig. \ref{Fig:3} (we suppose
that the vector $\mathbf{k}$ is located at the Fermi surface and $%
\varepsilon _{\mathbf{k+Q}}=0$ due to nesting). 
One can see that at small $\kappa $, i.e. in the vicinity of the Mott
transition one finds splitting of Hubbard subbands, while the QP remains
unsplit (Fig. 3a,b). In the narrow region of larger $\kappa $ the QP is
split in two peaks, and the splitting of the Hubbard subbands remain visible
(Fig. 3c). At intermediate values of $\kappa $ we find only splitting of the
QP peak, the two other peaks corresponding to the Hubbard subbands are
present (Fig. 3d,3e). 
Finally, in the weak coupling limit $\kappa =1$ we reproduce the two-peak
pseudogap, discussed in Refs.\ \onlinecite{TPSC,TPSCSE} (Fig. 3f). In a more
general case of $\gamma _{s,\mathbf{Q}}^{\nu ,0}\neq 1$ we expect a
pseudogap of the size $\sim \Delta (\gamma _{s,\mathbf{Q}0}^{\Delta
,0})^{1/2}$ in the weak coupling regime at small enough temperatures and
more complicated structures at strong $U$; see our numerical results below.

\section{Moriyaesque $\protect\lambda $ correction for the vertex}

\label{Sec:lambda}

The local approximation for the particle-hole irreducible vertex, considered
in Section II, is however not exact. In particular, the magnetic transition
temperature remains equal to its value in DMFT, and therefore it is
overestimated in both three- and two dimensions. In the latter case $T_{N}$
would remain finite, contrary to the Mermin-Wagner theorem.

In the D$\Gamma $A framework a reduction of $T_{N}$ would naturally arise
from a self-consistent solution of the D$\Gamma $A equations. An
alternative (simpler) way to fulfill the Mermin-Wagner theorem in 2D (and to
reduce the transition temperature in three dimensions) is to introduce a
correction to the susceptibility similar to the Moriya theory of weak
itinerant magnets\cite{Moriya}. To this end, we replace 
\begin{equation}
\chi _{\mathbf{q}\omega }^{s}\longrightarrow \left[ (\chi _{\mathbf{q}\omega
}^{s})^{-1}+\lambda _{\mathbf{q\omega }}\right] ^{-1}.  \label{his}
\end{equation}%
Formally the r.h.s. of Eq. (\ref{his}) is exact for some (unknown) $\lambda
_{\mathbf{q\omega }}$; in the following we assume $\lambda _{\mathbf{q\omega 
}}\simeq \lambda _{\mathbf{Q}0}\equiv \lambda $ since static fluctuations
with momentum $\mathbf{Q}$ predominate near the magnetic instability.
Instead of determining (as it was done in Moriya theory) $\lambda $ from the
fluctuation correction to the free energy, which is rather cumbersome in the
present approach, we (similar to TPSC) impose the fulfillment of the sumrule%
\begin{equation}
-\int_{-\infty }^{\infty }\frac{d\nu }{\pi }\mbox{Im}\Sigma _{\mathbf{k},\nu
}=U^{2}n(1-n/2)/2.  \label{As1}
\end{equation}%
This also implies%
\begin{equation}
\mbox{Re}\Sigma _{\mathbf{k},\nu }\simeq \frac{U^{2}n(1-n/2)}{2\nu }
\label{As2}
\end{equation}%
for $\nu \gg D$, 
\begin{figure}[t]
\includegraphics[width=8cm]{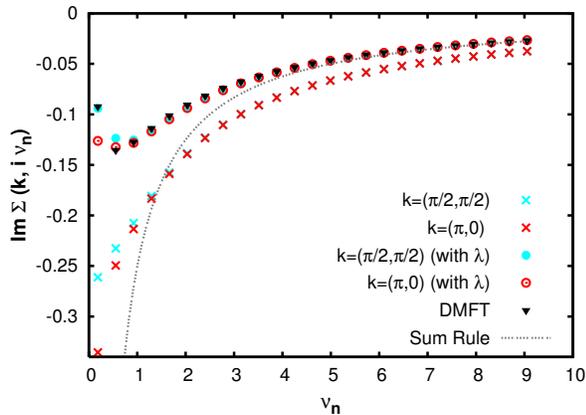}
\caption{(Color online) D$\Gamma $A self-energy on the Matsubara axis
calculated with and without Moriya $\protect\lambda$ correction for two
different points of the Fermi surface in the two dimensional Hubbard model
(at $U=D=4t$, $\protect\beta =1/T=17$); also shown is the DMFT self energy
for comparison. Notice that, without introducing the Moriya $\protect\lambda$
correction, one always observes a deviation of the high-frequency $\Sigma (%
\mathbf{k},i\protect\nu _{n})$ from the correct asymptotic behavior $\sim
U^2 n (1-\frac{n}{2})/(2 i \protect\nu_n) = U^{2}/(4i\protect\nu _{n})$
which is consistent with the self energy sum rule (see text).}
\label{FigSigwim}
\end{figure}
according to the Kramers-Kronig relation. The latter asymptotic behavior may
be very important to obtain the correct Fermi surface in the non-half-filled
case, but should be fulfilled also in the half-filled case to obtain correct
spectral functions. It is obviously violated in standard spin-fermion (also
paramagnon) approaches in two dimensions, where the N\'{e}el temperature ($%
T_N$) is finite without the $\lambda $ correction and the l.h.s. of Eqs. (%
\ref{As1}) and (\ref{As2}) are divergent at $T\longrightarrow T_{N}$.

The frequency dependence of the self-energy at the imaginary axis for the
two-dimensional Hubbard model ($U=D=4t$), calculated with and without $%
\lambda $ correction is compared in Fig. \ref{FigSigwim}. The $\lambda $
correction removes the divergence of the l.h.s. of Eqs. (\ref{As1}) and (\ref%
{As2}) at $T\rightarrow T_{N}^{\text{DMFT}}$ and leads to the correct
asymptotic behavior at large $\nu _{n}$. Without $\lambda$-correction (or,
alternatively, a self-consistent solution of the D$\Gamma $A equations) spin
fluctuations and their pertinent effect on the self energy are
overestimated. This is because the spin fluctuations result in a reduced
metalicity which in a second D$\Gamma $A iteration, i.e., the recalculation
of the local vertex with the less metallic Green function as an input\cite%
{dga_proc}, would reduce the spin fluctuations.

In two dimensions the sumrules (\ref{As1}) and (\ref{As2}) can be fulfilled
at all positive temperatures, and the actual transition temperature is zero,
as required by the Mermin-Wagner theorem.
As one can see from Eqs. (\ref{hiq}) and (\ref{SEAprox}), the
correlation length $\xi $ in two dimensions is exponentially divergent (with $\lambda $-correction): 
\[
\xi \propto \exp (b/T),
\]%
where the coefficient $b$ in the exponent is proportional to $U$. This is
evidently confirmed also by our numerical results shown in Fig. \ref%
{FigInvChiS}, where we have reported the values of the inverse of the spin
susceptibility at $\mathbf{Q}=(\pi ,\pi )$ calculated with the inclusion of
the $\lambda $-correction: The exponential divergence of $\xi $ for $%
T\rightarrow 0$ is directly reflected in an analogous behavior of the spin
susceptibility ($\chi _{\mathbf{{Q},0}}^{s}\sim A\xi ^{2}$, see Eq. (\ref%
{hiq})) at $T\rightarrow 0$.

In three dimensions, on the other hand, the sumrules (\ref{As1}) and (\ref%
{As2}) with $(\chi _{\mathbf{{Q},0}}^{s})^{-1}+\lambda >0$ can be fulfilled
only down to a certain temperature $T_{N}^{\text{D}\Gamma {\text{A}}}$,
which is reduced in comparison with $T_{N}^{\text{DMFT}}$ and determines the
phase transition temperature in the D$\Gamma $A approach.

\begin{figure}[t]
\includegraphics[width=8cm]{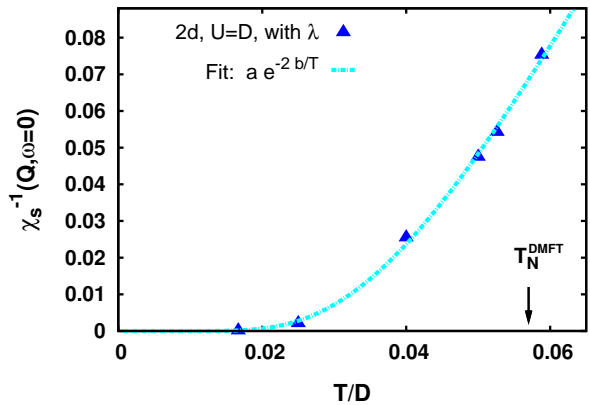}
\caption{Temperature dependence of the $\lambda$-corrected 
inverse antiferromagnetic susceptibility in two
dimensions (triangles) for $U=D=4t$. The data display an
exponential temperature dependence, consistent with the expected behavior of
$\xi$ (see text). The DMFT N\'{e}el temperature corresponding to this set of
parameters is marked with an arrow.}
\label{FigInvChiS}
\end{figure}

\section{Results for the Hubbard model in three dimensions}

\label{Sec:3D}

Let us turn to the results for the self-energy and spectral functions which
are obtained applying the Moriya $\lambda $ correction to the vertex of the D%
$\Gamma $A for the three dimensional system (the analytical continuation to
the real axis $i\nu _{n}\rightarrow \omega $ was done using the Pad\'{e}
algorithm). In this case, as mentioned above, the $\lambda $ correction is
expected to result in small -and only quantitative- changes of the final D$%
\Gamma $A results, because in $d=3$ (where the antiferromagnetic long-range
order survives at finite temperatures) the $\lambda $ correction produces
just a moderate reduction of the N\'{e}el temperature w.r.t. the DMFT value.

\begin{figure}[tb]
\includegraphics[width=8cm]{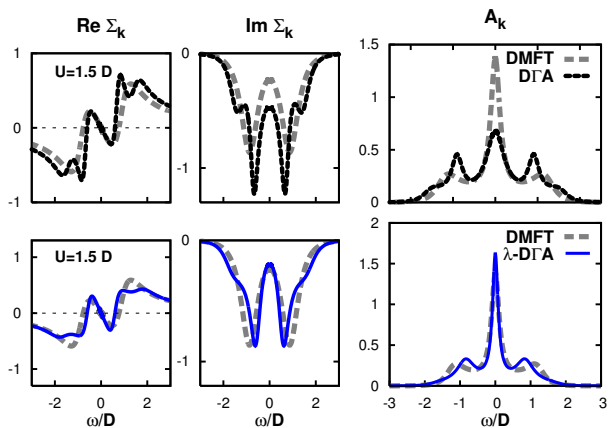}
\caption{(Color online) DMFT self-energies and spectral functions (grey dashed
 line) at $\mathbf{k}_F=(\protect\pi/2,\protect\pi/2,\protect\pi/2)$ for the Hubbard model in $d=3$  at $U= 1.5 D$ ($D=2\protect\sqrt{6}t$) and $\protect\beta =11.2$ (i.e., slightly above $T_N^{\mathrm{DMFT}}$) are compared with the corresponding D$\Gamma$A results with (lower row; solid blue line) and without (upper row; black dotted line) Moriya  $\protect \lambda$ correction. Note that (i) the non-local fluctuations modify only quantitatively the shape of the QP, but no pseudogap appears, and (ii) non-local correlation effects are further reduced by the inclusion of the Moriya $\protect\lambda$ correction.}
\label{Fig3d}
\end{figure}

\begin{figure}[tb]
\includegraphics[width=8cm]{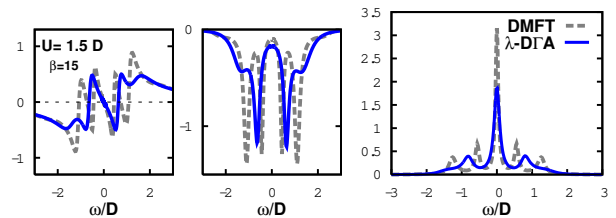}
\caption{(Color online) DMFT self-energies and  spectral functions at $\mathbf{k}_F=(\protect\pi /2,\protect\pi /2,\protect\pi/2)$ for the Hubbard model in 
$d=3$ at $U= 1.5 D$ ($D=2 \protect\sqrt{6}t$) and $\protect\beta =15$ compared with the corresponding D$\Gamma$A ones with $\protect\lambda$ correction. Lowering $%
T=1/\protect\beta$ towards $T_N^{D\Gamma A}$, qualitatively similar results
as without $\protect\lambda$ correction at higher $T$ (upper row of Fig.\ \protect\ref{Fig3d}, $\protect\beta=11.2$) are obtained.}
\label{Fig3db15}
\end{figure}

Our results, shown in Fig. \ref{Fig3d}, clearly confirm this expectation.
Specifically, we analyze the case, already considered in our previous study
Ref.\ \onlinecite{DGA1}, i.e., the three dimensional Hubbard model with $U=1.5$
(in the units of half the variance of the non-interacting DOS, being $%
D\equiv 2\sqrt{6}t$ for $d=3$), and $\beta =11.2$ (in units of $1/D$), which
corresponds to a temperature \textsl{slightly above} the DMFT N\'{e}el
temperature ($T_{N}^{\mathrm{DMFT}}$), but  appreciably higher than the
three-dimensional $T_{N}^{\text{D}\Gamma {\text{A}}}$ with $\lambda$
correction (an estimate of the $\lambda-$reduced  N\'{e}el temperature
gives $\beta^{\text{D}\Gamma \text{A}}=1/T_{N}^{D\Gamma A} \simeq 16.5$). In
this situation, as noticed in Ref.\ \onlinecite{DGA1} and shown in Fig.\ \ref{Fig3d} (first row), the standard D$\Gamma$A results display a sizable
renormalization of the quasiparticle (QP) peak present in the DMFT spectrum. However, no qualitative change in
the nature of the spectral functions can be observed. The inclusion of the
Moriya $\lambda $ correction, as shown in the second row of Fig. \ref{Fig3d}%
, reduces the renormalization effects due to non-local correlations: both
the real and the imaginary part of the D$\Gamma $A self-energy at low
frequency get very close to the DMFT values, and, obviously, the same
happens to the QP peak in $A(\mathbf{k},\omega )$. This result is easily
understood in terms of the reduction of $T_{N}$ determined by the Moriya
corrections, since the enhanced distance to the second-order
antiferromagnetic transition at $T_{N}$ leads to a reduction of the
spin-fluctuation and corrections to the DMFT self-energy. If we reduce the
temperature towards the D$\Gamma $A N\'{e}el temperature, antiferromagnetic
spin fluctuations become strong again, and as shown in Fig.\ \ref{Fig3db15},
we indeed find results which are qualitatively similar to those without $%
\lambda $ correction (first row of Fig.\ref{Fig3d}). In particular, in both
figures the quasiparticle weight is smaller in D$\Gamma $A than in DMFT in
agreement with the expected effect of antiferromagnetic fluctuations.

Summing up the results for the isotropic three dimensional system, we
emphasize that the principal consequence of the inclusion of the Moriya $%
\lambda $ correction is a shift of the region with appreciable non-local
correlation effects (i.e., the region where the D$\Gamma $A spectra
substantially differ from DMFT) to lower temperatures, i.e., to the
proximity of the \textquotedblleft new\textquotedblright\ line of the
antiferromagnetic phase transition. Our result demonstrates that for $d=3$
-with or without lambda correction- the extension of the region
characterized by relevant non-local correlations is relatively small even
for intermediate values of the interactions. This indicates, hence, that for 
$d=3$ DMFT represents indeed a good approximation, except for the region
close to the antiferromagnetic transition.

\section{Results for the Hubbard model in two dimensions}

\label{Sec:2D}

The effects of non-local correlations are -as one can imagine- much more
dramatic for a two-dimensional system. It is easy to figure out that the
divergence of the ladder diagrams in the spin channel leads to huge
non-local corrections in the D$\Gamma$A self-energy, which can differ also
qualitatively from the DMFT one. At the same time, one should expect that in
two dimensions the non-local correlation effects could be sensibly
overestimated by the D$\Gamma$A without the inclusion of the Moriya $\lambda$
correction. As we have discussed in Section \ref{Sec:lambda}, these
corrections are essential to fulfill the Mermin-Wagner theorem, pushing the
N\'eel temperature from the DMFT value down to zero. Hence, for any finite
temperature the antiferromagnetic fluctuations are reduced. The effects of
the divergence of the spin ladder diagrams are also to some extent
attenuated in the formula for the D$\Gamma$A self-energy, because of the
extra dimension gained at $T=0$ due to the transformation of the Matsubara
summation to a frequency integral on the r.h.s. of Eq. (\ref{SE}).

\begin{figure}[t]
{\includegraphics[width=85mm]{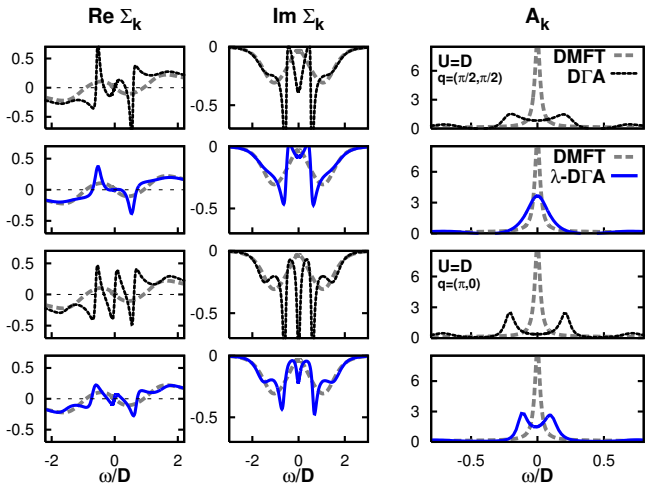}}
\caption{(Color online) D${\Gamma}$A results for the half-filled
two-dimensional Hubbard model at $U=D=4t $, $\protect\beta=17$ (just
slightly above $T_N^{\mathrm{DMFT}}$) computed without  (first and third row; black dotted line) and with (second and fourth row; blue solid line) the Moriya $\protect\lambda$ correction, and compared with DMFT (grey dashed line). The D$\Gamma$A calculations in the non-selfconsistent scheme show a
clear pseudogap opening for the $\mathbf{k}$ points of the non-interacting
FS, but more pronounced in the antinodal direction. Within the $\protect%
\lambda$-corrected scheme, one can still notice a pseudogap opening but only
in the antinodal direction, while in the nodal direction a strongly damped
QP appears. }
\label{Fig2Db17}
\end{figure}

\begin{figure}[t]
{\includegraphics[width=8cm]{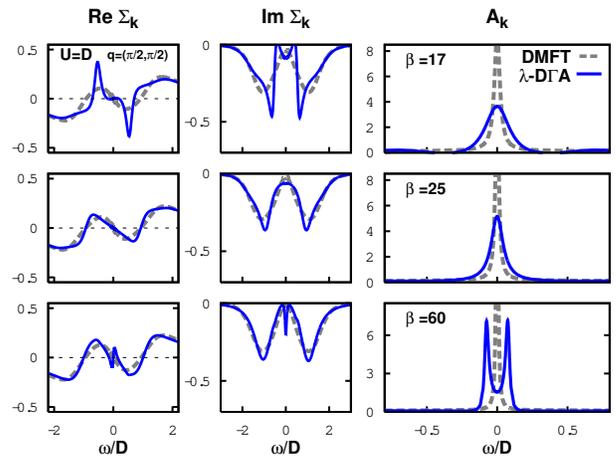}}
\caption{(Color online) Temperature evolution of the D${\Gamma}$A results
for the half-filled two-dimensional Hubbard model at $U=D=4t$ in the nodal
direction, computed with the Moriya $\protect\lambda$ correction, and
compared with DMFT. A clear pseudogap emerges at the lowest temperature ($%
\protect\beta=60$), similarly to the results of the non-self consistent
scheme in the proximity of $T_N^{\mathrm{DMFT}}$ (see Fig. \protect\ref%
{Fig2Db17}).}
\label{Fig2Dnod}
\end{figure}

\begin{figure}[tbh]
{\includegraphics[width=8cm]{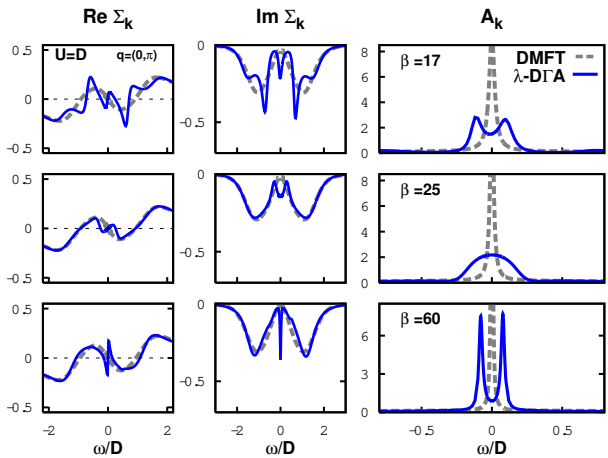}}
\caption{(Color online) Same as in Fig. \protect\ref{Fig2Dnod} in the
antinodal direction. As expected, a very pronounced pseudogap characterizes
the lowest temperature results. The behavior of the spectral functions is
not completely monotonous, as the pseudogap seems to disappear at $\protect%
\beta=25$. At all temperatures, however, the pseudogap features are always
more marked in the antinodal than in the nodal direction.}
\label{Fig2Dant}
\end{figure}

In the light of these considerations, we can more easily interpret the
results of the D$\Gamma$A for the two-dimensional Hubbard model, which are
presented in Figs. \ref{Fig2Db17}, \ref{Fig2Dnod}, \ref{Fig2Dant}.
Specifically, we start the analysis of the two dimensional case, by
evaluating the effects of the Moriya $\lambda$ correction for the D$\Gamma$A
results computed for the half-filled Hubbard model with $U=4t$ at a
temperature ($\beta=17$) \textsl{slightly above} the corresponding $T_N$ in
DMFT.

In the first/third row of Fig.\ \ref{Fig2Db17}, we show the D$\Gamma $A
self-energy and spectral function at the Fermi surface (FS) at the nodal [$%
\mathbf{q}=(\frac{\pi }{2},\frac{\pi }{2})$]/antinodal [$\mathbf{q}=(\pi ,0)$%
] points computed without Moriya correction. One can clearly observe that,
in contrast to the three dimensional case, the D$\Gamma $A spectra
qualitatively differ from the original DMFT one because (i) a pseudogap
appears at low frequencies and (ii) the spectra are markedly anisotropic in
the nodal/antinodal direction, as the observed pseudogap is evidently more
pronounced at the antinodal points.

As discussed above, the Moriya $\lambda$ correction is however expected to
be much more important in the two-dimensional than in the three dimensional case . This is confirmed by the
results shown in the second/fourth row of Fig.\ \ref{Fig2Db17}. In these
panels the reduction of the non-local effects due to the inclusion of the
Moriya $\lambda$ correction in D$\Gamma$A is evident. It is important
noticing, however, that although the distance in the phase-diagram from the
actual antiferromagnetic transition (occurring at $T=0$ for $d=2$) is
considerably larger than for $d=3$, non-local correlation effects are
nonetheless extremely strong. Turning to the details, we still observe a
remarkable anisotropy in the D$\Gamma$A spectra after the inclusion of
Moriya correction, with a strongly renormalized QP peak in the nodal
direction and a rather clear pseudogap in the antinodal direction. The
results of D$\Gamma$A (implemented with the Moriya correction)
indicate, hence, that for the two-dimensional system at half-filling
antiferromagnetic fluctuation effects predominate in a wide region of the
phase diagram, determining the onset of an anisotropic pseudogap in the
spectra also for considerably high temperatures, qualitatively similar to
that observed in underdoped cuprates.\cite{pseudogapCuprates}

The inclusion of the Moriya correction in D$\Gamma $A allows us to extend
our analysis to the low-temperature regime $T<T_{N}^{\text{DMFT}}$. In
particular, we are interested to study the evolution of the spectral
function when the temperature is considerably reduced compared to the DMFT
value $T_{N}^{\mathrm{DMFT}}$. In Figs. \ref{Fig2Dnod} and \ref{Fig2Dant} we
report the D$\Gamma $A calculation for the self-energy and the spectral
function for the same case considered above ($U=4t$, half-filling) for three
different decreasing temperatures ($\beta =17$, shown already in Fig. \ref%
{Fig2Db17}, $\beta =25$ and $\beta =60$) in the nodal and antinodal
direction, respectively. First, we note that the anisotropy in the
self-energy and the spectra remains visible at all temperatures. In
addition, a marked tendency towards a completely gapped spectrum can be seen
at the lowest temperature: At lowest temperature ($\beta =60$) a pseudogap
appears also in the nodal direction, while the pseudogap already present in
the antinodal direction becomes remarkably more profound.  At this
temperature, therefore, the anisotropy is reduced in comparison to the
higher T cases- due to the strong depletion of spectral weight at $\omega =0$. 
This results can be understood in terms of the closer proximity to the
antiferromagnetic instability at $T=0$, and is consistent with the marked
pseudogap visible in the $k$-$integrated$ spectral function obtained by
means of cluster DMFT in Ref.\ \onlinecite{clusterDMFT}.

\vskip 5mm

It is worth noticing, however, that the temperature evolution towards the
formation of a fully gapped spectrum at $T\rightarrow 0$ does not appear to
be completely monotonous. The effects of the non-local fluctuations seems to
be slightly weaker in the D$\Gamma$A results for $\beta=25$ (second row in
Figs. \ref{Fig2Dnod}-\ref{Fig2Dant}), than for $\beta=17$ (first row). More
specifically, this is visible in the slightly \textsl{more Fermi-liquid-like}
behavior of the real and imaginary part of the self-energies at $\beta=25$
in comparison to $\beta=17$. 

\begin{figure}[t]
{\includegraphics[width=85mm]{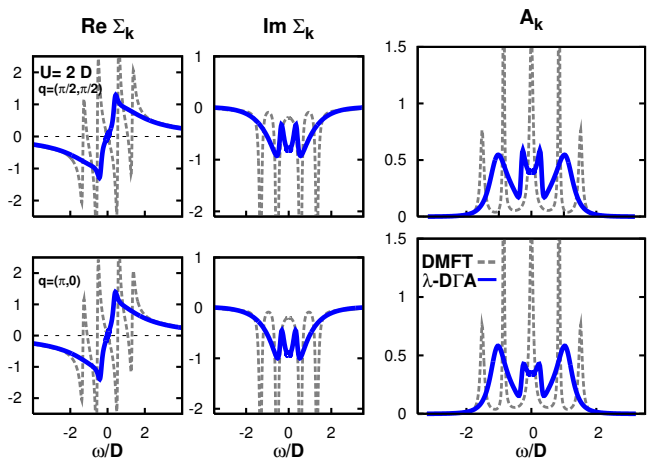}}
\caption{Color online) D${\Gamma}$A results with $\protect\lambda$
corrections for the half-filled two-dimensional Hubbard model at $U=2D=8t $, 
$\protect\beta=40$ compared with the corresponding DMFT ones.}
\label{Fig2Db40}
\end{figure}

A possible interpretation of this specific feature of our results is to
relate the non-monotonous temperature evolution in the D$\Gamma $A spectral
function to a competition between non-local and local mechanisms capable of
destroying coherent excitations: (i) The (non-local) antiferromagnetic
fluctuations, which become less pronounced with increasing $T$, making the system more metallic  ($\chi _{\mathbf{{Q},0}}^{s}=$ $8.9\cdot 10^{3}$, $39.26$,
and $13.28$, for $\beta =60$, $25$, and $17$, respectively); and at the
same time (ii) the thermal loss of coherence, which is at the origin of so-called crossover region in the (purely local) DMFT and reflects increasing values 
of the quasiparticle damping ($\gamma =-$Im$%
\Sigma (0)=0.009,0.021,0.034,$ for the three considered temperatures
respectively) . The relevance of the interplay between these two
mechanisms is an interesting issue raised by our D$\Gamma $A results. It
might also be related to a similar non-monotonous trend in the cluster DMFT
phase diagram reported by Park \emph{et al.} \cite{Park08}.

The D$\Gamma $A results at stronger interaction ($U=2D$ and $\beta=40$) are
presented in Fig. \ref{Fig2Db40}. At the considered low temperature the
local DMFT spectral functions have peaky structure, because we solve the
impurity problem of DMFT by means of exact diagonalization (ED), which
treats only finite number of sites. Note, however, the D$\Gamma$A spectral
functions are continuous due to momenta- and frequency sums in the Eq. (\ref%
{SE0}), even though ED is employed as an impurity solver. The nonlocal
spectral functions show the splitting of the quasiparticle peak due to
magnetic correlations, which is similar to the structure (d) in Fig. \ref%
{Fig:3} discussed in Sect. IV \cite{Kusunose}. Closer to the Mott transition
(i.e. at even stronger U) we also expect the formation of the structures
(a)-(c) of Fig. \ref{Fig:3}.

The presented results demonstrate that the D$\Gamma $A approach -with the
inclusion of the Moriya corrections- allows for a non trivial analysis of
the effects of long-range spatial correlations in every region of the phase
diagrams of strongly interacting fermionic systems both in two and three
dimensions.

\section{$\mathbf{k}$-resolved spectral functions in two dimensions}

\label{Sec:ARPES}

Let us now calculate the $\mathbf{k}$-dependence of the spectral functions
in the directions of high symmetry, as can be observed in angular resolved
photoemission spectroscopy (ARPES). It is worthwhile remarking that, in
contrast to the cluster extensions of DMFT, this does not require any kind
of interpolation in $\mathbf{k}$-space: Due to the diagrammatic nature of
the D$\Gamma$A, the spectra for every chosen $\mathbf{k}$ point in the first
Brillouin zone are easily computed via Eq. (\ref{SE}).

\begin{figure}[tb]
{\includegraphics[width=8.cm,angle=0]{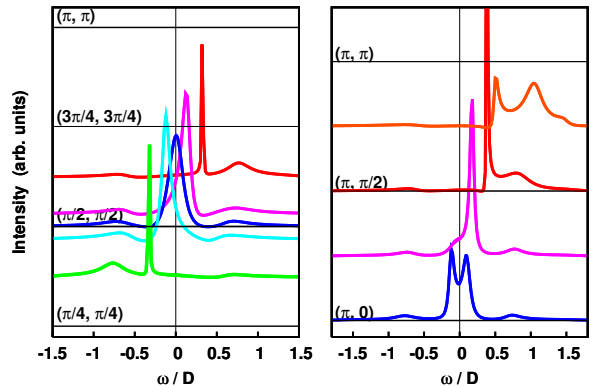}}
\caption{(Color online) $\mathbf{k}$-resolved D$\Gamma$A spectra along the
nodal (left) and antinodal (right) direction for the half-filled
two-dimensional Hubbard model at $U=D=4t$ and $\protect\beta=17$, calculated
with the Moriya $\protect\lambda$ correction.}
\label{Fig:2DkL}
\end{figure}

Here, in Fig. \ref{Fig:2DkL},  we present D$\Gamma$A results with Moriya $\lambda$ correction for the
same case previously considered in Fig. \ref{Fig2Db17} (second and fourth
row). As it is often done, we consider two different $\mathbf{k}$-paths
along the Brillouin zone, the first one along the nodal direction [$(0,0)
\rightarrow (\pi,\pi)$, left panel] and the second one right at the border
of the Brillouin zone, crossing the antinodal point at the FS [$(\pi,\pi)
\rightarrow (\pi,0)$, right panel].

Our analysis of the $\mathbf{k}$-resolved D$\Gamma$A results allows us to
appreciate the evolution of the main features of the D$\Gamma$A spectral
functions. Specifically, we observe that for the points most far away from
the FS, the spectral functions display similar features in the two cases: A
relatively narrow peak separated from a broader maximum at higher energies,
which represents the incoherent processes building up the (upper) Hubbard
band. When proceeding in the direction of the FS, as expected, the narrow
peak moves towards the Fermi energy, while the broad feature becomes less
pronounced. A qualitative difference between the two selected paths emerges
only in the vicinity of the FS: The shift of the narrow peak down to zero
energy is frozen along the second path, consistent with the opening of the
anisotropic pseudogap in the antinodal direction, while it continues to
shift down to the Fermi level in the nodal directions.

It is also worth noticing the occurrence in both cases of a slight
broadening of the narrow peak while approaching the FS. This trend, which is
markedly different from any FL expectation, could be understood in terms of
the maximum of Im $\Sigma(\mathbf{{k},\omega)}$ appearing at zero frequency
(see again Fig. \ref{Fig2Db17}) for both directions. The enhanced value of Im%
$\Sigma(\mathbf{{k},\omega)}$ at low frequencies, which is ultimately
responsible for the opening of the pseudogap starting at the antinodal
points, determines a loss of coherence and, hence, the observed broadening
of the peak, while it moves closer to the FS.

\vskip 15mm

\section{Conclusion}

\label{Sec:Conclusion}

Based on the representation of the nonlocal self energy which considers the
effect of the bare Coulomb interaction and charge (spin) fluctuations, we
have extended the recently introduced dynamical vertex approximation (D$%
\Gamma$A) by including a Moriyaesque $\lambda$ correction to the local
vertex in Section \ref{Sec:lambda}. The value of $\lambda$ is determined
from the sum rule which relates $\omega$-integrated self energy and
occupation and allows for a proper reduction of the DMFT N\'eel temperature,
in two dimensions even to $T_N=0$ so that the Mermin-Wagner theorem is
fulfilled. This correction is therefore particularly important for two
dimensions, where spin fluctuations are especially strong. Without the
Moriya $\lambda$ correction, a much more involved self-consistent solution
of the D$\Gamma$A equations would be necessary to yield similar results.

The method we have introduced here allows for a treatment of non local
long-range spatial correlation in finite dimensional systems. In three
dimensions, pronounced effects of non-local spin fluctuations are found only
close to the antiferromagnetic phase transition. This is in contrast to the
two dimensional case where antiferromagnetic fluctuations completely
reshuffle the spectrum, also far away from the antiferromagnetic phase
transition at $T_N=0$, leading eventually to the formation of a pseudogap.
Qualitatively, the spectral functions can be understood by means of the
analytical formula for the self energy proposed in Section \ref{Sec:approx}.
Calculating several D$\Gamma$A self energies along the high symmetry lines
of the Brillouin zone, we obtain the momentum dependence of the spectral
functions, which could be directly compared with the ARPES data.

D$\Gamma$A can serve as a very promising method for future studies of the
Hubbard model at non-integer filling, in particular in the vicinity of the
antiferromagnetic quantum critical point.\cite{SpFerm1,QCP} A further
important development would be also the generalization of the method to the
multi-orbital case, to analyze the effects of non local correlations beyond
DMFT in realistic bandstructure calculations.\cite{LDADMFT}

\textit{Acknowledgments.} We thank W.\ Metzner, M. Capone, C.\ Castellani,
G.\ Sangiovanni, and R.\ Arita for stimulating discussions and are indebted
to M. Capone also for providing the DMFT(ED) code which has served as a
starting point.  This work has been supported by the EU-Indian cooperative network MONAMI. The work of AK was supported by the Russian Basic Research
Foundation through Grants No.~1941.2008.2 (Support of Scientific Schools)
and 07-02-01264a.

\end{document}